\documentclass[a4paper]{article}
\usepackage{mathptmx,graphicx,amsmath,amsfonts,amsthm,amssymb,url,cite,hyperref}
\usepackage{amsmath,amssymb,amsfonts}
\usepackage{algorithm,booktabs,algpseudocode}
\usepackage{xcolor,authblk}

\newtheorem{theorem}{Theorem}

\newtheorem{lemma}{Lemma}

\algnotext{EndFor}
\algnotext{EndIf}
\algnotext{EndWhile}

\newcommand{\vc}[1]{\mathbf{#1}}
\newcommand{\argmin}{\operatorname{argmin}}
\newcommand{\Ret}{\textbf{return }}
\newcommand{\To}{\textbf{to }}

\newtheorem{heuristic}{Heuristic}

\DeclareMathOperator{\Search}{\operatorname{search}}

\begin{document}

\title{Solving the Shortest Vector Problem in Lattices Faster Using Quantum Search}

\author[1]{Thijs Laarhoven}
\author[2]{Michele Mosca}
\author[3]{Joop van de Pol}

\affil[1]{Dept. of Mathematics and Computer Science, Eindhoven Univ. Tech.\protect\\
P.O. Box 513, 5600 MB Eindhoven, The Netherlands\protect\\
\url{t.m.m.laarhoven@tue.nl}}

\affil[2]{Institute for Quantum Computing and Dept. of C\&O, Univ. Waterloo\protect\\
200 Univ. Avenue West, Waterloo, ON, N2L 3G1, Canada\protect\\
and \protect\\
Perimeter Institute for Theoretical Physics\protect\\31 Caroline Street North, Waterloo, ON, N2L 2Y5, Canada \protect\\
\url{michele.mosca@uwaterloo.ca}}

\affil[3]{Dept. of Computer Science, University of Bristol\protect\\
Merchant Venturers Building, Woodland Road, Bristol, BS8 1UB, UK\protect\\
\url{joop.vandepol@bristol.ac.uk}}

\date{}

\maketitle

\begin{abstract}
By applying Grover's quantum search algorithm to the lattice algorithms of Micciancio and Voulgaris, Nguyen and Vidick, Wang et al., and Pujol and Stehl\'{e}, we obtain improved asymptotic quantum results for solving the shortest vector problem. With quantum computers we can provably find a shortest vector in time $2^{1.799n + o(n)}$, improving upon the classical time complexity of $2^{2.465n + o(n)}$ of Pujol and Stehl\'{e} and the $2^{2n + o(n)}$ of Micciancio and Voulgaris, while heuristically we expect to find a shortest vector in time $2^{0.312n + o(n)}$, improving upon the classical time complexity of $2^{0.384n + o(n)}$ of Wang et al. These quantum complexities will be an important guide for the selection of parameters for post-quantum cryptosystems based on the hardness of the shortest vector problem.
\end{abstract}


\section{Introduction}
\label{sec:intro}

Large-scale quantum computers will redefine the landscape of computationally secure cryptography, including breaking public-key cryptography based on integer factorization or the discrete logarithm problem \cite{shor97} or the Principle Ideal Problem in in real quadratic number fields \cite{hallgren07}, providing sub-exponential attacks for some systems based on elliptic curve isogenies \cite{childs10b}, speeding up exhaustive searching \cite{grover96, boyer98} and (with appropriate assumptions about the computing architecture) finding collisions and claws \cite{brassard98, buhrman05, ambainis04}, among many other quantum algorithmic speed-ups \cite{childs10, smith12, mosca09}.

Currently, a small set of systems \cite{bernstein08} are being studied intensely as possible systems to replace those broken by large-scale quantum computers. These systems can be implemented with conventional technologies and to date seem resistant to substantial quantum attacks. It is critical that these systems receive intense scrutiny for possible quantum or classical attacks. This will boost confidence in the resistance of these systems to (quantum) attacks, and allow us to fine-tune secure choices of parameters in practical implementations of these systems. 

One such set of systems bases its security on the computational hardness of certain lattice problems. Since the late 1990s, there has been a lot of research into the area of lattice-based cryptography, resulting in encryption schemes~\cite{hoffstein98,regev05}, digital signature schemes~\cite{gentry08,lyubashevsky12} and even fully homomorphic encryption schemes~\cite{gentry09,brakerski12}. Each of the lattice problems that underpin the security of these systems can be reduced to the shortest vector problem. For a more detailed summary on the security of lattice-based cryptography, see~\cite{laarhoven12,pol11}.

In this paper, we closely study the best-known algorithms for solving the shortest vector problem on a lattice, and how quantum algorithms may speed up these attacks. By challenging and improving the best asymptotic complexity of such attacks, we increase the confidence in the security of lattice-based schemes. Understanding these attacks is critical when selecting key-sizes and other security parameters. 


\subsection{Lattices}
\label{sub:lat}

Lattices are discrete subgroups of $\mathbb{R}^n$. Given a set of $n$ linearly independent vectors $B = \{\vc{b}_1, \ldots, \vc{b}_n\}$ in $\mathbb{R}^n$, we define the lattice generated by these vectors as $L = \left\{\sum_{i = 1}^n \lambda_i \vc{b}_i: \lambda_i \in \mathbb{Z}\right\}$. We call the set $B$ a basis of the lattice $L$. This basis is not unique; applying a unimodular matrix transformation to the vectors of $B$ leads to a new basis $B'$ of the same lattice $L$.

In lattices, we generally work with the Euclidean or $\ell_2$-norm, which we will denote by $\|\cdot\|$. For bases $B$, we write $\|B\| = \max_i \|\vc{b}_i\|$. We refer to a vector $\vc{s} \in L \setminus \{\vc{0}\}$ such that $\|\vc{s}\| \leq \|\vc{v}\|$ for any $\vc{v} \in L \setminus \{\vc{0}\}$ as a shortest vector of the lattice. Its length is denoted by $\lambda_1(L)$. Given a basis $B$, we write $\mathcal{P}(B) = \left\{\sum_{i = 1}^n \lambda_i \vc{b}_i: 0 \leq \lambda_i < 1\right\}$ for the fundamental domain of $B$.

One of the most important hard problems in the theory of lattices is the Shortest Vector Problem (SVP). Given a basis of a lattice, the Shortest Vector Problem consists of finding a shortest vector in this lattice. In many applications, finding a short vector instead of a shortest vector is also sufficient. The Approximate Shortest Vector Problem with approximation factor $\gamma$ (SVP$_{\gamma}$) asks to find a non-zero lattice vector $\vc{v} \in L$ with length bounded from above by $\|\vc{v}\| \leq \gamma \lambda_1(L)$.


\subsection{Related work}
\label{sub:relwork}

The Approximate Shortest Vector problem is integral in the cryptanalysis of lattice-based cryptography \cite{gama08}. For small values of $\gamma$, this problem is known to be NP-hard \cite{ajtai98,khot05}, while for certain exponentially large $\gamma$, polynomial time algorithms exist, such as the LLL algorithm of Lenstra, Lenstra and Lov\'{a}sz~\cite{lenstra82}. Other algorithms trade running time for a better approximation factor $\gamma$, such as the LLL algorithm with deep insertions \cite{schnorr94} and the BKZ algorithm of Schnorr and Euchner \cite{schnorr94}. The latter algorithm requires an exact SVP algorithm for lower dimensions as a subroutine. The current state-of-the-art for classically finding short vectors is BKZ 2.0~\cite{chen11}, which is essentially the original BKZ algorithm with the improved SVP subroutine of Gama et al.~\cite{gama10}. Implementations of this algorithm, due to Chen and Nguyen~\cite{chen11}, and Aono and Naganuma~\cite{aono12}, currently dominate the Lattice Challenge Hall of Fame~\cite{latticechallenge12}.

In 2003, Ludwig~\cite{ludwig03} used quantum algorithms to speed up one such basis reduction algorithm, Random Sampling Reduction (RSR), which is due to Schnorr~\cite{schnorr03}. By replacing a random sampling from a big list by a quantum search, Ludwig achieves a quantum algorithm that is asymptotically faster than previous results. Ludwig also details the effect that this faster quantum algorithm would have had on the practical security of the lattice-based encryption scheme NTRU~\cite{hoffstein98}, had there been a quantum computer in 2005.

In the cryptanalysis of schemes that are based on lattice problems, it is often sufficient to find a short vector and not necessarily a shortest vector. In this setting, basis reduction algorithms such as BKZ seem to be more efficient than exact (and generally exponential) SVP algorithms. However, SVP solvers are still relevant for lattice-based cryptography, because the BKZ algorithm also requires an efficient low-dimensional SVP algorithm as a subroutine. Several methods are known for finding a shortest vector and in theory each of these could be used as a subroutine for BKZ. For SVP solvers there is a similar online challenge~\cite{svpchallenge12}, where the record is currently held by Kuo et al.~\cite{kuo11}.

\subsubsection{Enumeration.} The classical method for finding shortest vectors is enumeration, dating back to work by Pohst~\cite{pohst81}, Kannan~\cite{kannan83} and Fincke and Pohst~\cite{fincke85} in the first half of the 1980s. In order to find a shortest vector, one enumerates all lattice vectors inside a giant ball around the origin. If the input basis is only LLL-reduced, enumeration runs in $2^{O(n^2)}$ time, where $n$ is the lattice dimension. The algorithm by Kannan uses a stronger preprocessing of the input basis, and runs in $2^{O(n \log n)}$ time. 
Both approaches use only polynomial space in $n$.

\subsubsection{Sieving/Saturation.} In 2001, Ajtai et al.~\cite{ajtai01} introduced a technique called sieving, leading to the first algorithm to solve SVP in time $2^{O(n)}$. Starting with a huge list of short vectors, the algorithm repeatedly applies a sieve to this list to end up with a smaller list of shorter lattice vectors. After several iterations we hope to be left with a list of lattice vectors of length $O(\lambda_1(L))$. Due to the size of the list, the space requirement of sieving is $2^{O(n)}$. Later work~\cite{hanrot11,micciancio10b,regev04,nguyen08} investigated the constants in both exponents and ways to reduce these.

Recently, in 2009, Micciancio and Voulgaris~\cite{micciancio10b} started a new branch of sieving algorithms, which may be more appropriately called saturation algorithms. While sieving starts out with a long list and repeatedly applies a sieve to reduce its length, saturation algorithms iteratively add vectors to an initially empty list, hoping that at some point the space of short lattice vectors is ``saturated'', and two of the vectors in the list are at most $\lambda_1(L)$ apart. The time and space requirements of these algorithms are also $2^{O(n)}$. In 2009, Pujol and Stehl\'{e}~\cite{pujol09} showed that with this method, SVP can provably be solved in time $2^{2.465n + o(n)}$.

\subsubsection{Voronoi.} In 2010, Micciancio and Voulgaris presented another algorithm for solving SVP based on constructing the Voronoi cell of the lattice \cite{micciancio10}. In time $2^{2n + o(n)}$ and space $2^{n + o(n)}$, this algorithm is able to find a shortest vector in any lattice. Currently this is the best provable asymptotic result for classical SVP solvers.

\subsubsection{Practice.} While many methods have surpassed the enumeration algorithms in terms of classical provable asymptotic time complexities, in practice the enumeration methods still dominate the field. The version of enumeration that is currently used in practice is due to Schnorr and Euchner~\cite{schnorr94} with improvements by Gama et al.~\cite{gama10}. It does not incorporate the stronger version of preprocessing of Kannan~\cite{kannan83} and hence has an asymptotic time complexity of $2^{O(n^2)}$. However, due to the small hidden constants in the exponents and the exponential space complexity of the other algorithms, enumeration is actually faster than other methods for common values of $n$. That said, the other methods are still quite new, so a further study of these other methods may tip the balance.


\subsection{Quantum search}
\label{sub:qa}

In this paper we will study how quantum algorithms can be used to speed up the SVP algorithms outlined above. For this, we will make use of Grover's quantum search algorithm~\cite{grover96}, which considers the following problem:

Given a list $L$ of length $N$ and a function $f: L \to \{0,1\}$, such that the number of elements $e \in L$ with $f(e) = 1$ is small. Construct an algorithm ``$\Search$'' that, given $L$ and $f$ as input, returns an $e \in L$ with $f(e) = 1$, or determines that (with high probability) no such $e$ exists. We assume for simplicity that $f$ can be evaluated in unit time.

\subsubsection{Classical algorithm.} With classical computers, the natural way to find such an element is to go through the whole list, until one of these elements is found. This takes on average $O(N)$ time. This is also optimal up to a constant factor; no classical algorithm can find such an element in less than $\Omega(N)$ time.

\subsubsection{Quantum algorithm.} Using quantum search \cite{grover96, boyer98}, we can find such an element in time $O(\sqrt{N})$. This is optimal up to a constant factor, as any quantum algorithm needs at least $\Omega(\sqrt{N})$ evaluations of $f$ \cite{bennett97}. \\ 

Throughout the paper, we will write $x \leftarrow \Search_{e \in L}(f(e) = 1)$ to highlight subroutines that perform a search in a long list. This assignment returns true if an element $e \in L$ with $f(e) = 1$ exists (and assigns such an element to $x$), and returns false if no such $e$ exists. This allows us to give one description for both the classical and quantum versions of each algorithm, as the only difference between the two versions is which version of the subroutine is used.

For both of these classical and quantum algorithms, we assume a RAM model of computation where the $j$th entry of the list $L$ can be looked up in constant time (or polylogarithmic time). In the case that $L$ is a virtual list where the $j$th element can be computed in time polynomial in the length of $j$ (thus polylogarithmic in the length of the list $L$), then look-up time is not an issue. When $L$ is indeed an unstructured list of values, for classical computation, the assumption of a RAM-like model has usually been valid in practice. However, there are fundamental reasons for questioning it \cite{bernstein09}, and there are practical computing architectures where the assumption does not apply.
In the case of quantum computation, a practical RAM-like quantum memory looks particularly challenging, especially for first generation quantum computers. Some authors have studied the limitations of quantum algorithms in this context \cite{grover04, bernstein09, jeffery11}.

Some algorithms (e.g. \cite{ambainis04}) must store a large database of information in regular quantum memory (that is, memory capable of storing quantum superpositions of states). In contrast, quantum searching an actual list of $N$ (classical) strings requires the $N$ values to be stored in quantumly addressable classical memory (e.g. as Kuperberg discusses in \cite{kuperberg11}) and $O(\log N)$ regular qubits.  Quantumly addressable classical memory in principle could be much easier to realize in practice than regular qubits. Furthermore, quantum searching for a value $x \in \{0,1\}^n$ satisfying $f(x) =1$ for a function $f: \{0,1\}^n \rightarrow \{0,1\}$ and which can be implemented by a circuit on $O(n)$ qubits only requires $O(n)$ regular qubits, and there is no actual list to be stored in memory. In this paper, the quantum search algorithms used require the lists of size $N$ to be stored in quantumly addressable classical memory and use $O(\log N)$ regular qubits and $O(\sqrt{N})$ queries into the list of numbers.

In this work, we consider (conventional) classical RAM memories for the classical algorithms, and RAM-like quantumly addressable classical memories for the quantum search algorithms. This is both a first step for future studies in assessing the impact of more practical quantum architectures, and also represents a more conservative approach in determining parameter choices for lattice-based cryptography that should be resistant against the potential power of quantum algorithmic attacks.


\subsection{Contributions and outline}
\label{sub:cont}

In this paper, we show that quantum algorithms can significantly speed up sieving and saturation algorithms. The constant in the exponent decreases by approximately $25\%$ in all cases, leading to an improvement upon both provable and heuristic asymptotic results for solving the Shortest Vector Problem:
\begin{itemize}
  \item Provably, we can find a shortest vector in any lattice in time $2^{1.799n + o(n)}$.
  \item Heuristically, we can find a shortest vector in any lattice in time $2^{0.312n + o(n)}$.
  \item Extrapolating from classical experiments, with quantum computers we expect to be able to find a shortest vector in any lattice in time about $2^{0.39n}$.
\end{itemize}
Table~\ref{tab:th} contains a comparison between our contributions and previous results, in both the classical and quantum setting. While the Voronoi Cell algorithm is asymptotically the best algorithm in the provable classical setting, our quantum saturation algorithm has better asymptotics in the provable quantum setting.

Why do we only consider sieving and saturation algorithms, and not the more practical enumeration or the theoretically faster Voronoi cell algorithms? It turns out that it is not as simple to significantly speed up these algorithms using similar techniques. For some intuition why this is the case, see Appendix~\ref{sec:other}.

The outline of this paper is as follows. In Section~\ref{sec:sieving} we look at sieving algorithms, and how quantum algorithms lead to speed-ups. In Section~\ref{sec:saturation}, we look at saturation algorithms, and their estimated time and space complexities on a quantum computer. Technical details regarding some of these results can be found in Appendices~\ref{app:nvs} and \ref{app:ps}. 

\begin{table}[t]
\centering
\caption{A comparison of the results as expressed in logarithmic leading order terms.}\label{tab:th}
\begin{tabular}{p{4.22cm}rrrrl}
\toprule
 & \multicolumn{2}{r}{\textbf{Classical}} & \multicolumn{2}{r}{\textbf{Quantum}} & \\
\textbf{Algorithm} & Time & Space & Time & Space & \\
\midrule
(Enumeration) 									& $O(n \log n)$ & $O(1)$ & - & - & \, (App.~\ref{sec:other}) \\
Pujol and Stehl\'{e} \cite{pujol09}	 			& $2.47n$ & $1.24n$ & $1.80n$ & $1.29n$ & \, (Sec.~\ref{sec:lsb}) \\
(Voronoi)                                    	& $2.00n$ & $1.00n$ & - & - & \, (App.~\ref{sec:other}) \\
\midrule
Micciancio and Voulgaris \cite{micciancio10b} 	& $0.52n$ & $0.21n$ & $0.39n$ & $0.21n$ & \, (Sec.~\ref{sec:gs}) \\
Nguyen and Vidick \cite{nguyen08}				& $0.42n$ & $0.21n$ & $0.32n$ & $0.21n$ & \, (Sec.~\ref{sec:nvs})\\
Wang et al. \cite{wang11}						& $0.39n$ & $0.26n$ & $0.32n$ & $0.21n$ & \, (Sec.~\ref{sec:ws}) \\
\bottomrule
\end{tabular}
\end{table}


\section{Sieving algorithms}
\label{sec:sieving}

Sieving was first introduced by Ajtai et al.~\cite{ajtai01} and later improved theoretically \cite{regev04,nguyen08,micciancio10b,hanrot11} and practically \cite{nguyen08,wang11} in various papers. In these algorithms, first an exponentially long list of lattice vectors is generated. Then, by iteratively applying a sieve to this list, the size of the list, as well as the lengths of the vectors in the list are reduced. After a polynomial number of applications of the sieve, we hope to be left with a short but non-empty list of very short vectors, from which we can then obtain a shortest vector of the lattice.





\subsection{The Heuristic Algorithm of Nguyen and Vidick}
\label{sec:nvs}


\begin{algorithm}[t]
\caption{The Heuristic Sieve Algorithm of Nguyen and Vidick}
\label{alg:nvs}
\begin{algorithmic}[1]
\Require{An LLL-reduced basis $B$ of $L$, and constants $\gamma \in (\frac{2}{3}, 1)$ and $N = 2^{O(n)}$}
\Ensure{A short non-zero lattice vector $\vc{s}$}
\State $S \leftarrow \emptyset$
\For{$i \leftarrow 1$ \textbf{to} $N$}
	\State $\vc{v} \in_R B_n(\vc{0}, \|B\|) \cap L$
	\State $S \leftarrow S \cup \{\vc{v}\}$
\EndFor
\While{$S \setminus \{\vc{0}\} \neq \emptyset$}
	\State $S_{\text{prev}} \leftarrow S \setminus \{\vc{0}\}$
	\State $R \leftarrow \max_{\vc{v} \in S_{\text{prev}}} \|\vc{v}\|$
	\State $C \leftarrow \{\vc{0}\}$
	\State $S \leftarrow \emptyset$
	\ForAll{$\vc{v} \in S_{\text{prev}}$}
		\If{$\vc{c} \leftarrow \Search_{\vc{c} \in C}(\|\vc{v} - \vc{c}\| \leq \gamma R)$} \label{alg:nvs:search1}
			\State $S \leftarrow S \cup \{\vc{v} - \vc{c}\}$
		\Else
			\State $C \leftarrow C \cup \{\vc{v}\}$	
		\EndIf
	\EndFor
\EndWhile
\State $\vc{s} \leftarrow \argmin_{\vc{v} \in S_{\text{prev}}} \|\vc{v}\|$ \label{alg:nvs:search2}
\State \Ret $\vc{s}$
\end{algorithmic}
\end{algorithm}


Nguyen and Vidick \cite{nguyen08} considered a heuristic, practical variant of the sieve algorithm of Ajtai et al.~\cite{ajtai01}, which provably returns a shortest vector under a certain natural, heuristic assumption. A slightly modified but equivalent version of this algorithm is given in Algorithm~\ref{alg:nvs}.

\subsubsection{Description of the algorithm.} The algorithm starts by generating a big list $S$ of random lattice vectors with length at most $\|B\|$. Then, by repeatedly applying a sieve to this list, shorter lists of shorter vectors are obtained, until the list is completely depleted. In that case, we go back one step, and look for the closest pair of lattice vectors in the last non-empty list.

The sieving step consists of splitting the previous list $S_{\text{prev}}$ in a set of `centers' $C$ and a new list of vectors $S$ that will be used for the next sieve. For each vector $\vc{v}$ in $S_{\text{prev}}$, the algorithm first checks if a vector $\vc{c}$ in $C$ exists that is close to $\vc{v}$. If this is the case, then we add the difference $\vc{v} - \vc{c}$ to $S_{\text{prev}}$. If this is not the case, then $\vc{v}$ is added to $C$. Since the set $C$ consists of vectors with a bounded norm and a specified minimum distance between any two points, one can bound the size of $C$ from above using a result of Kabatiansky and Levenshtein~\cite{kabatiansky78} regarding sphere packings. In other words, $C$ will be sufficiently small, so that the list $S$ will be sufficiently large. After applying the sieve, we discard all vectors in $C$ and apply the sieve again to the vectors in $S_{\text{prev}} = S$.

At each iteration of the sieve, the maximum norm of the vectors in the list decreases from some constant $R$ to at most $\gamma R$, where $\gamma$ is some geometric factor smaller than $1$. Nguyen and Vidick conjecture that throughout the algorithm, the longest vectors in $S$ are uniformly distributed over the space of all $n$-dimensional vectors with norms between $\gamma R$ and $R$.

\begin{heuristic} \cite{nguyen08} \label{heur:nv}
At any stage of Algorithm~\ref{alg:nvs}, the vectors in $S \cap C_n(\gamma R, R)$ are uniformly distributed in $C_n(\gamma R, R)$, where $C_n(r_1, r_2) = \{\vc{x} \in \mathbb{R}^n: r_1 \leq \|\vc{x}\| \leq r_2\}$.
\end{heuristic}

\subsubsection{Classical complexities.} In Line~\ref{alg:nvs:search1} of Algorithm~\ref{alg:nvs}, we have highlighted an application of a search subroutine that could be replaced by a quantum search. Using a standard classical search algorithm for this subroutine, under this heuristic assumption Nguyen and Vidick give the following estimate for the time and space complexity of their algorithm.

\begin{lemma} \cite{nguyen08}
On a classical computer, assuming that Heuristic~\ref{heur:nv} holds, Algorithm~\ref{alg:nvs} will return a shortest vector of a lattice in time at most $2^{0.415n + o(n)}$ and space at most $2^{0.208n + o(n)}$.
\end{lemma}

\subsubsection{Quantum complexities.} If we use a quantum search subroutine in Line~\ref{alg:nvs:search1}, the complexity of this subroutine decreases from $O(|C|)$ to $O(\sqrt{|C|})$. Since this search is part of the bottleneck for the time complexity, applying a quantum search here will decrease the running time significantly. Note that in Line~\ref{alg:nvs:search2}, it also seems like a search of a list is performed. In reality, this final search of $S_{\text{prev}}$ can be done in constant time by using appropriate data structures, e.g., by keeping the vectors in $S$ and $S_{\text{prev}}$ sorted from short to long, or by manually keeping track of the shortest vector in $S$.

Since replacing the classical search by a quantum search does not change the internal behaviour of the algorithm, the estimates and heuristics are as valid as they were in the classical setting. The time complexity does change, as the following theorem explains. For details, see Appendix~\ref{app:nvs}.

\begin{theorem} \label{thm:nvs}
On a quantum computer, assuming that Heuristic~\ref{heur:nv} holds, Algorithm~\ref{alg:nvs} will return a shortest vector of a lattice in time $2^{0.312n + o(n)}$ and space $2^{0.208n + o(n)}$.
\end{theorem}

In other words, applying quantum search to Nguyen and Vidick's sieve algorithm leads to a $25\%$ decrease in the exponent of the runtime.


\subsection{The Heuristic Algorithm of Wang et al.}
\label{sec:ws}

To improve upon the time complexity of the algorithm of Nguyen and Vidick, Wang et al.~\cite{wang11} introduced a further trade-off between the time complexity and the space complexity. Their algorithm uses two lists of centers $C_1$ and $C_2$ and two geometric factors $\gamma_1$ and $\gamma_2$, instead of the single list $C$ and single geometric factor $\gamma$ in the algorithm of Nguyen and Vidick. For details, see \cite{wang11}.

\subsubsection{Classical complexities.} The classical time complexity of this algorithm is bounded from above by $\tilde{O}(|S| \cdot (|C_1| + |C_2|))$, while the space required is at most $O(|S| + |C_1| + |C_2|)$. Optimizing the constants $\gamma_1$ and $\gamma_2$ leads to $\gamma_1 = 1.0927$ and $\gamma_2 \to 1$, with an asymptotic time complexity of less than $2^{0.384n + o(n)}$ and a space complexity of about $2^{0.256n + o(n)}$.

\subsubsection{Quantum complexities.} By using the quantum search algorithm for searching the lists $C_1$ and $C_2$, the time complexity is reduced to $\tilde{O}(|S| \cdot (\sqrt{|C_1|} + \sqrt{|C_2|}))$, while the space complexity remains $O(|S| + |C_1| + |C_2|)$. Re-optimizing the constants for a minimum time complexity leads to $\gamma_1 \to \sqrt{2}$ and $\gamma_2 \to 1$, leading to the same time and space complexities as the quantum-version of the algorithm of Nguyen and Vidick. Due to the simpler algorithm and smaller constants, a quantum version of the algorithm of Nguyen and Vidick will most likely be more efficient than a quantum version of the algorithm of Wang et al.



\section{Saturation algorithms}
\label{sec:saturation}

Saturation algorithms were only recently introduced by Micciancio and Voulgaris~\cite{micciancio10b}, and further studied by Pujol and Stehl\'{e}~\cite{pujol09} and Schneider~\cite{schneider11}. Instead of starting with a huge list and making the list smaller and smaller, this method starts with a small or empty list, and keeps adding more and more vectors to the list. Building upon the same result of Kabatiansky and Levenshtein about sphere packings~\cite{kabatiansky78}, we know that if the list reaches a certain size and all vectors have a norm bounded by a sufficiently small constant, two of the vectors in the list must be close to one another. Thus, if we can guarantee that new short lattice vectors keep getting added to the list, then at some point, with high probability, we can find a shortest vector as the difference between two of the list vectors.



\subsection{The Provable Algorithm of Pujol and Stehl\'{e}}
\label{sec:lsb}


\begin{algorithm}[t]
\caption{The Provable Saturation Algorithm of Pujol and Stehl\'{e}}\label{alg:lsb}
\begin{algorithmic}[1]
\Require{An LLL-reduced basis $B$ of $L$, and constants $\mu \simeq \lambda_1(L)$, $\xi > \frac{1}{2}$, $R > 2\xi$}
\Ensure{A non-zero lattice vector $\vc{s}$ of norm less than $\mu$}
\State $\gamma \leftarrow 1 - \frac{1}{n}$
\State $T \leftarrow \emptyset$
\State $N_1 \in_R [0, N_1^{\max} - 1]$
\For{$i \leftarrow 1$ \To $N_1$}
	\State $\vc{x} \in_R B_n(\vc{0}, \xi \mu)$
	\State $\vc{v}' \leftarrow \vc{x} \mod \mathcal{P}(B)$
	\While{$\vc{t} \leftarrow \Search_{\vc{t} \in T}(\|\vc{v}' - \vc{t}\| < \gamma\|\vc{v}'\|)$} \label{alg:lsb:search1}
		\State $\vc{v}'  \leftarrow \vc{v}' - \vc{t}$
	\EndWhile
	\State $\vc{v} \leftarrow \vc{v}' - \vc{x}$
	\If{$\|\vc{v}\| \geq R \mu$}
		\State $T \leftarrow T \cup \{\vc{v}\}$
	\EndIf
\EndFor
\State $S \leftarrow \emptyset$
\For{$i \leftarrow 1$ \To $N_2$}
	\State $\vc{x} \in_R B_n(\vc{0}, \xi \mu)$
	\State $\vc{v}' \leftarrow \vc{x} \mod \mathcal{P}(B)$
	\While{$\vc{t} \leftarrow \Search_{\vc{t} \in T}(\|\vc{v}' - \vc{t}\| < \gamma\|\vc{v}'\|)$} \label{alg:lsb:search2}
		\State $\vc{v}' \leftarrow \vc{v}' - \vc{t}$
	\EndWhile
	\State $\vc{v} \leftarrow \vc{v}' - \vc{x}$
	\State $S \leftarrow S \cup \{\vc{v}\}$\
\EndFor
\State $\{\vc{s}_1, \vc{s}_2\} \leftarrow \Search_{\{\vc{s}_1, \vc{s}_2\} \in S \times S}(0 < \|\vc{s}_1 - \vc{s}_2\| < \mu)$ \label{alg:lsb:search3}
\State \Ret $\vc{s}_1 - \vc{s}_2$
\end{algorithmic}
\end{algorithm}


Using the Birthday paradox, Pujol and Stehl\'{e}~\cite{pujol09} showed that the constant in the exponent of the time complexity of the original algorithm of Micciancio and Voulgaris~\cite[Section 3.1]{micciancio10b} can be reduced by almost $25\%$. The algorithm is presented in Algorithm~\ref{alg:lsb}.

\subsubsection{Description of the algorithm.} The algorithm can roughly be divided in three stages, as follows.

First, the algorithm generates a long list $T$ of lattice vectors with norms between $R \mu$ and $\|B\|$. This `dummy' list is only used for technical reasons, and in practice one does not seem to need such a list. Note that besides the actual lattice vectors $\vc{v}$, to generate this list we also consider slightly perturbed vectors $\vc{v}'$ which are not in the lattice, but are at most $r \mu$ away from $\vc{v}$. This is purely a technical modification to make the proofs work, as experiments show that without such perturbed vectors, saturation algorithms also work fine \cite{micciancio10, pujol09, schneider11}.

After generating $T$, we generate a fresh list of short lattice vectors $S$. The procedure for generating these vectors is similar to that of generating $T$, with two exceptions: (i) now all sampled lattice vectors are added to $S$ (regardless of their norms), and (ii) the vectors are reduced with the dummy list $T$ rather than with vectors in $S$. The latter guarantees that the vectors in $S$ are i.i.d.

Finally, when $S$ has been generated, we hope that it contains two distinct lattice vectors $\vc{s}_1$, $\vc{s}_2$ that are at most $\mu$ apart. So we search $S \times S$ for a pair $\{\vc{s}_1, \vc{s}_2\}$ of close, distinct lattice vectors, and return their difference.

\subsubsection{Classical complexities.} With a classical search applied to the subroutines in Lines~\ref{alg:lsb:search1}, \ref{alg:lsb:search2}, and \ref{alg:lsb:search3}, Pujol and Stehl\'{e} obtained the following results.

\begin{lemma} \cite{pujol09}
Let $\xi \approx 0.9476$ and $R \approx 3.0169$. Then, using polynomially many queries to Algorithm~\ref{alg:lsb}, we can find a shortest vector in a lattice with probability exponentially close to $1$, using time at most $2^{2.465n + o(n)}$ and space at most $2^{1.233n + o(n)}$.
\end{lemma}

\subsubsection{Quantum complexities.} Applying a quantum search algorithm to the search-subroutines in Lines~\ref{alg:lsb:search1}, \ref{alg:lsb:search2}, and \ref{alg:lsb:search3} leads to the following result. Details are given in Appendix~\ref{app:ps}.

\begin{theorem} \label{thm:lsb}
Let $\xi \approx 0.9086$ and $R \approx 3.1376$. Then, using polynomially many queries to the quantum version of Algorithm~\ref{alg:lsb}, we can find a shortest vector in a lattice with probability exponentially close to $1$, using time at most $2^{1.799n + o(n)}$ and space at most $2^{1.286n + o(n)}$.
\end{theorem}

So the constant in the exponent of the time complexity decreases by about $27\%$ when using quantum search.

\paragraph{Remark.} If we generate $S$ in parallel, we can potentially achieve a time complexity of 
$2^{1.470n + o(n)}$, by setting $\xi \approx 1.0610$ and $R \approx 4.5166$. However, it would require exponentially many parallel quantum computers of size $O(n)$ to achieve a substantial theoretical speed-up over the $2^{1.799n + o(n)}$ of Theorem~\ref{thm:lsb}.
(Recall that quantum searching a list of $c^n$ elements (with $c > 1$) requires the list to be stored in quantumly addressable classical memory (versus regular quantum memory) and otherwise can be searched using only $O(n)$ qubits and $O(c^{n/2})$ queries to the list.)

\subsection{The Heuristic Algorithm of Micciancio and Voulgaris}
\label{sec:gs}


\begin{algorithm}[t]
\caption{The Heuristic Saturation Algorithm of Micciancio and Voulgaris}\label{alg:gs}
\begin{algorithmic}[1]
\Require{An LLL-reduced basis $B$ of $L$, and a constant $C_0$}
\Ensure{A short non-zero lattice vector $\vc{s}$}
\State $S \leftarrow \{\vc{0}\}$
\State $Q \leftarrow \emptyset$
\State $c \leftarrow 0$
\While{$c < C_0$}
	\If{$Q \neq \emptyset$}
		\State $\vc{v} \in_R Q$
		\State $Q \leftarrow Q \setminus \{\vc{v}\}$
	\Else
		\State $\vc{v} \in_R B_n(\vc{0}, \|B\|) \cap L$
	\EndIf
	\While{$\vc{s} \leftarrow \Search_{\vc{s} \in S}(\max\{\|\vc{s}\|, \|\vc{v} - \vc{s}\|\} \leq \|\vc{v}\|)$} \label{alg:gs:search1}
		\State $\vc{v} \leftarrow \vc{v} - \vc{s}$
	\EndWhile
	\While{$\vc{s} \leftarrow \Search_{\vc{s} \in S}(\max\{\|\vc{v}\|, \|\vc{v} - \vc{s}\|\} \leq \|\vc{s}\|)$} \label{alg:gs:search2}
		\State $S \leftarrow S \setminus \{\vc{s}\}$
		\State $Q \leftarrow Q \cup \{\vc{v} - \vc{s}\}$
	\EndWhile
	\If{$\vc{v} = \vc{0}$}
		\State $c \leftarrow c + 1$
	\Else
		\State $S \leftarrow S \cup \{\vc{v}\}$
	\EndIf
\EndWhile
\State $\vc{s} \leftarrow \argmin_{\vc{v} \in S \setminus \{\vc{0}\}} \|\vc{v}\|$
\State \Ret $\vc{s}$
\end{algorithmic}
\end{algorithm}

In practice, just like sieving algorithms, saturation algorithms are much faster than their worst-case running times and provable time complexities suggest. Micciancio and Voulgaris \cite{micciancio10b} gave a heuristic variant of their saturation algorithm, for which they could not give a (heuristic) bound on the time complexity, but with a better bound on the space complexity, and a better practical time complexity. The algorithm is given in Algorithm~\ref{alg:gs}.

\subsubsection{Description of the algorithm.} The algorithm is similar to Algorithm~\ref{alg:lsb}, with the following main differences: (i) we do not explicitly generate two lists $S$, $T$ to apply the birthday paradox; (ii) we do not use the geometric factor $\gamma < 1$ but always reduce a vector if it can be reduced; (iii) we also reduce the existing list vectors with newly sampled vectors, so that each two vectors in the list are pairwise Gauss-reduced; and (iv) instead of specifying the number of iterations, we run the algorithm until we reach a predefined number of collisions $C_0$.

\subsubsection{Classical complexities.} Micciancio and Voulgaris state that the algorithm above has an experimental time complexity of about $2^{0.52n}$ and a space complexity which is most likely bounded from above by $2^{0.208n}$ due to the kissing constant~\cite[Section 5]{micciancio10b}. This is much faster than the theoretical time complexity of $2^{1.799n}$ of the quantum-enhanced saturation algorithm discussed in Section~\ref{sec:lsb}.

\paragraph{Remark 1.} In practice, the algorithm of Micciancio and Voulgaris is faster than the one of Nguyen and Vidick of Section~\ref{sec:nvs}, even though the leading term in the exponent is larger. So asymptotically, this algorithm is dominated by the algorithm of Nguyen and Vidick, but in practice and for small dimensions, the algorithm of Micciancio and Voulgaris seems to perform better.

\paragraph{Remark 2.} Schneider states \cite{schneider11} that the time complexity scales like $2^{0.57n - 23.5}$, instead of the $2^{0.52n}$ claimed by Micciancio and Voulgaris. Although asymptotically this time complexity is worse than the one of Micciancio and Voulgaris, the cross-over point of these rough approximations is around $n \approx 470$. So for most values of $n$ that SVP solvers handle in practice, the term $-23.5$ is more significant than the small increase caused by $n$, and the conjectured time complexity of Schneider is better than that of Micciancio and Voulgaris.

\subsubsection{Quantum complexities.} To this heuristic algorithm, the quantum speed-ups can also be applied. Generally, these saturation algorithms generate a list $S$ of reasonably short lattice vectors by (i) first sampling a long, random lattice vector $\vc{v} \in L$; (ii) reducing the vector $\vc{v}$ with lattice vectors already in $S$; (iii) possibly reducing the vectors in $S$ with this new vector $\vc{v}$; and (iv) finally adding $\vc{v}$ to $S$. The total classical time complexity of these algorithms is of the order $|S|^2$ due to (ii) and (iii), but by applying quantum speed-ups to these steps, this becomes $|S|^{3/2}$. This means that the exponent in the time complexity is generally reduced by about $25\%$, which is comparable to the improvement in Section~\ref{sec:lsb}. In practice, we therefore expect a time complexity of about $2^{0.39n}$ for the heuristic algorithm of Micciancio and Voulgaris with quantum search speed-ups, with constants that may make this algorithm faster than the sieving algorithm of Section~\ref{sec:nvs}.

\subsubsection*{Acknowledgments.}

This report is partly a result of fruitful discussions at the Lorentz Center Workshop on Post-Quantum Cryptography and Quantum Algorithms, Nov. 5--9, Leiden, The Netherlands. In particular, we would like to thank Felix Fontein, Nadia Heninger, Stacey Jeffery, Stephen Jordan, Michael Schneider, Damien Stehl\'{e} and Benne de Weger for the valuable discussions there.

The first author is supported by DIAMANT and ECRYPT II (ICT-2007-216676). The second author is supported by Canada's NSERC (Discovery, FREQUENCY, and CREATE CryptoWorks21), MPrime, CIFAR, ORF and CFI; IQC and Perimeter Institute are supported in part by the Government of Canada and the Province of Ontario. The third author is supported in part by EPSRC via grant EP/I03126X.


\appendix


\section{Analysis of the Sieve Algorithm of Nguyen and Vidick}
\label{app:nvs}

Nguyen and Vidick showed that if their heuristic assumption holds, the time and space complexities of their algorithm can be bounded from above as follows.

\begin{lemma} \cite{nguyen08}
On a classical computer, assuming Heuristic~\ref{heur:nv} holds, Algorithm~\ref{alg:nvs} will return a shortest vector of a lattice in time $2^{2 c_h n + o(n)}$ and space $2^{c_h n + o(n)}$, where $\frac{2}{3} < \gamma < 1$ and
\begin{align}
c_h = -\log_2(\gamma) - \frac{1}{2} \log_2\left(1 - \frac{\gamma^2}{4}\right).
\end{align}
\end{lemma}

To obtain a minimum time complexity, $\gamma$ should be chosen as close to $1$ as possible. Letting $\gamma \to 1$ leads to an asymptotic time complexity of less than $2^{0.415n + o(n)}$ and an asymptotic space complexity of less than $2^{0.208n + o(n)}$.

To obtain these estimates, it is first noted that the sizes of $S$ and $C$ are bounded from above by $2^{c_h n + o(n)}$. The space complexity is therefore bounded from above by $O(|S| + |C|) = 2^{c_h n + o(n)}$, and since for every element in $S$ the algorithm has to search the list $C$, the time complexity is bounded from above by $\tilde{O}(|S| \cdot |C|) = 2^{2c_h n + o(n)}$.

Using Grover's algorithm for searching the list $C$, the time complexity decreases to $\tilde{O}(|S| \cdot \sqrt{|C|}) = 2^{\frac{3}{2} c_h n + o(n)}$, while the space complexity remains the same. This leads to the following result.

\begin{lemma}
On a quantum computer, assuming Heuristic~\ref{heur:nv} holds, Algorithm~\ref{alg:nvs} will return a shortest vector of a lattice in time $2^{\frac{3}{2} c_h n + o(n)}$ and space $2^{c_h n + o(n)}$.
\end{lemma}

Optimizing $\gamma$ to obtain a minimum time complexity again corresponds to letting $\gamma$ tend to $1$ from below, leading to an asymptotic time complexity of $2^{0.312n + o(n)}$ and space complexity of $2^{0.208n + o(n)}$, as stated in Theorem~\ref{thm:nvs}.


\section{Analysis of the Saturation Algorithm of Pujol and Stehl\'{e}}
\label{app:ps}

In the classical setting, the time complexities of the different parts of the algorithm are as follows. The constants are explained in the lemma below.
\begin{itemize}
  \item Cost of generating $T$: $\tilde{O}(N_1^{\max} \cdot |T|) = 2^{(c_g + 2c_t)n + o(n)}$.
  \item Cost of generating $S$: $\tilde{O}(N_2 \cdot |T|) = 2^{(c_g + c_b/2 + c_t)n + o(n)}$.
  \item Cost of searching $S$ for a pair of close vectors: $\tilde{O}(|S|^2) = 2^{(2c_g + c_b)n + o(n)}$.
\end{itemize}
The space complexity is at most $O(|T| + |S|) = 2^{\max(c_t, c_g + c_b/2)n + o(n)}$. This leads to the following lemma.
%
%
\begin{lemma} \cite{pujol09}
Let $\xi > \frac{1}{2}$ and $R > 2\xi$, and suppose $\mu > \lambda_1(L)$. Then, with $c_b$, $c_t$, $c_g$, $N_B$, $N_V$, $N_G$, $N_1^{\max}$, $N_2$ chosen according to:
\begin{align}
c_b &= \log_2(R) + 0.401, & N_B &= 2^{c_b n + o(n)}, \label{eq:1} \\
c_t &= \frac{1}{2}\log_2\left(1 + \frac{\xi}{R - 2\xi}\right) + 0.401, & N_T &= 2^{c_t n + o(n)}, \\
c_g &= \frac{1}{2}\log_2\left(\frac{4\xi^2}{4\xi^2 - 1}\right),& N_G &= 2^{c_g n + o(n)}, \\
N_1^{\max} &= 2^{(c_g + c_t)n + o(n)}, & N_2 &= 2^{(c_g + c_b/2)n + o(n)}, \label{eq:4}
\end{align}
with probability at least $\frac{1}{16}$, Algorithm~\ref{alg:lsb} returns a lattice vector $\vc{s} \in L \setminus \{\vc{0}\}$ with $\|\vc{s}\| < \mu$, in time at most $2^{tn + o(n)}$ and space at most $2^{sn + o(n)}$, where $t$ and $s$ are given by
\begin{align}
t = \max\left(c_g + 2c_t, c_g + \frac{c_b}{2} + c_t, 2c_g + c_b\right), \quad s = \max\left(c_t, c_g + \frac{c_b}{2}\right).
\end{align}
\end{lemma}
In the quantum setting, the costs are as follows.
\begin{itemize}
  \item Cost of generating $T$: $\tilde{O}(N_1^{\max} \cdot \sqrt{|T|}) = 2^{(c_g + 3c_t/2)n + o(n)}$.
  \item Cost of generating $S$: $\tilde{O}(N_2 \cdot \sqrt{|T|}) = 2^{(c_g + c_b/2 + c_t/2)n + o(n)}$.
  \item Cost of searching $S$ for a pair of close vectors: $\tilde{O}(\sqrt{|S|^2}) = 2^{(c_g + c_b/2)n + o(n)}$.
\end{itemize}
The total space complexity is still the same as in the classical setting, i.e., at most $O(|T| + |S|) = 2^{\max(c_t, c_g + c_b/2)n + o(n)}$. This leads to the following lemma.

\begin{lemma}
Let $\xi > \frac{1}{2}$ and $R > 2\xi$, and suppose $\mu > \lambda_1(L)$. Then, with $c_b$, $c_t$, $c_g$, $N_B$, $N_V$, $N_G$, $N_1^{\max}$, $N_2$ chosen according to Equations~\eqref{eq:1} to \eqref{eq:4}, with probability at least $\frac{1}{16}$, Algorithm~\ref{alg:lsb} returns a lattice vector $\vc{s} \in L \setminus \{\vc{0}\}$ with $\|\vc{s}\| < \mu$ on a quantum computer in time at most $2^{\tilde{t}n + o(n)}$ and space at most $2^{\tilde{s}n + o(n)}$, where $\tilde{t}$ and $\tilde{s}$ are given by
\begin{align}
\tilde{t} = \max\left(c_g + \frac{3c_t}{2}, c_g + \frac{c_b}{2} + \frac{c_t}{2}, c_g + \frac{c_b}{2}\right), \quad \tilde{s} = \max\left(c_t, c_g + \frac{c_b}{2}\right).
\end{align}
\end{lemma}

Optimizing $\xi$ and $R$ for the minimum time complexity, we get $\xi \approx 0.9086$ and $R \approx 3.1376$ as in Theorem~\ref{thm:lsb}. Note that if $S$ is generated in parallel with exponentially many quantum computers, the cost of the second part of the algorithm becomes negligible, and the exponent in the time complexity changes to
\begin{align}
\tilde{t}' = \max\left(c_g + \frac{3c_t}{2}, c_g + \frac{c_b}{2}\right).
\end{align}
In that case, the optimal choice of $\xi$ and $R$ (with respect to minimizing the time complexity) would be $\xi \approx 1.0610$ and $R \approx 4.5166$, leading to a time complexity of less than $2^{1.470n + o(n)}$.


\section{Other SVP algorithms}
\label{sec:other}

\subsection{Enumeration}
Recall that enumeration considers all lattice vectors inside a giant ball around the origin that is known to contain at least one lattice vector. Let $L$ be a lattice with basis $\{\vc{b}_1,\ldots,\vc{b}_n\}$. Consider each lattice vector $\vc{u} \in L$ as a linear combination of the basis vectors, i.e., $\vc{u} = \sum_i u_i \vc{b}_i$. Now, we can represent each lattice vector by its coefficient vector $(u_1,\ldots,u_n)$. We would like to have all combinations of values for $(u_1,\ldots,u_n)$ such that the corresponding vector $\vc{u}$ lies in the ball. We could try any combination and see if it lies within the ball by computing the norm of the corresponding vector, but there is a smarter way that ensures we only consider vectors that lie within the ball and none that lie outside.

To this end, enumeration algorithms search from right to left, by identifying all values for $u_n$ such that there might exist $u_1',\ldots,u_{n-1}'$ such that the vector corresponding to $(u_1',\ldots,u_{n-1}',u_n)$ lies in the ball. To identify these values $u_1',\ldots,u_{n-1}'$, enumeration algorithms use the Gram-Schmidt orthogonalization of the lattice basis as well as the projection of lattice vectors. Then, for each of these possible values for $u_n$, the enumeration algorithm considers all possible values for $u_{n-1}$ and repeats the process until it reaches possible values for $u_1$. This leads to a search which is serial in nature, as each value of $u_n$ will lead to different possible values for $u_{n-1}$ and so forth. Unfortunately, we can only really apply the quantum search algorithm to problems where the list of objects to be searched is known in advance.

One might suggest to forego the smart way to find short vectors and just search all combinations of $(u_1,\ldots,u_n)$ with appropriate upper and lower bounds on the different $u_i$'s. Then it becomes possible to apply quantum search, since we now have a predetermined list of vectors and just need to compute the norm of each vector. However, it is doubtful that this will result in a faster algorithm, because the recent heuristic changes by Gama et al.~\cite{gama10} have reduced the running time of enumeration dramatically (roughly by a factor $2^{n/2}$) and these changes only complicate the search area further by changing the ball to an ellipsoid. There seems to be no simple way to apply quantum search to the enumeration algorithms that are currently used in practice, but perhaps the algorithms can be modified in some way.

\subsection{Voronoi cell}
Consider a set of points in the Euclidean space. For any given point in this set, its Voronoi cell is the region that contains all vectors that lie closer to this point than to any of the other points in the set. Now, given a Voronoi cell, we define a relevant vector to be any vector in the set whose removal from the set will change this particular Voronoi cell. If we pick our lattice as the set and we consider the Voronoi cell around the zero vector, then any shortest vector is also a relevant vector. Furthermore, given the relevant vectors of the Voronoi cell we can solve the closest vector problem in $2^{2n + o(n)}$ time.

So how can we compute the relevant vectors of the Voronoi cell of a lattice $L$? Micciancio and Voulgaris~\cite{micciancio10} show that this can be done by solving $2^n - 1$ instances of CVP in the lattice 2$L$.
However, in order to solve CVP we would need the relevant vectors which means we are back to our original problem. However, Micciancio and Voulgaris show that these instances of CVP can also be solved by solving several related CVP instances in a lattice of lower rank. They give a basic and an optimized version of the algorithm. The basic version only uses LLL as preprocessing and solves all these related CVP instances in the lower rank lattice separately. As a consequence, the basic algorithm runs in time $2^{3.5n + o(n)}$ and in space $2^{n + o(n)}$. The optimized algorithm uses a stronger preprocessing for the lattice basis, which takes exponential time. But since the most expensive part is the computation of the Voronoi relevant vectors, this extra preprocessing time does not increase the asymptotic running time. In fact, having the reduced basis decreases the asymptotic running time to $\tilde{O}(2^{3n})$. Furthermore, the optimized algorithm employs a trick that allows it to reduce $2^k$ CVP instances in a lattice of rank $k$ to a single instance of an enumeration problem related to the same lattice. The optimized algorithm solves CVP in time $\tilde{O}(2^{2n})$ using $\tilde{O}(2^n)$ space.

Now, in the basic algorithm, it would be possible to speed up the routine that solves the CVP given the Voronoi relevant vectors using a quantum computer. It would also be possible to speed up the routine that removes non-relevant vectors from the list of relevant vectors using a quantum computer. Combining these two changes gives a quantum algorithm with an asymptotic running time $\tilde{O}(2^{2.5n})$, which is still slower than the optimized classical algorithm. It is not possible to apply these same speedups to the optimized algorithm due to the aforementioned trick with the enumeration problem. The algorithm to solve this enumeration problem makes use of a priority queue, which means the search is not trivially parallellized. Once again, there does not seem to be a simple way to apply quantum search to this special enumeration algorithm. However, it may be possible that the algorithm can be modified in such a way that quantum search can be applied.

\end{document}